\documentclass[aps,showpacs]{revtex4}
\usepackage{times}
\usepackage{graphicx}
\begin{document}
\bibliographystyle{apsrev}
\title{Exciton-optical-phonon coupling: comparison with experiment for ZnO quantum wells}
\author{T.~Makino}
\email[Electronic mail: ]{tmakino@riken.jp}
\author{Y. Segawa}
\affiliation{Photodynamics Research Center, RIKEN (The Institute of Physical and Chemical Research), Aramaki-aza-Aoba 519-1399, Sendai 980-0845, Japan}
\author{M.~Kawasaki}
\altaffiliation[Also at: ]{Combinatorial Materials Exploration and Technology, Tsukuba 
305-0044, Japan}
\affiliation{Institute for Materials Research, Tohoku University,
Sendai 980-8577, Japan}

\date{\today}

\begin{abstract}
The temperature-dependent linewidths of excitons in ZnO quantum wells were studied by measuring absorption spectra from 5~K to room temperature. We deduced experimentally the exciton-longitudinal-optical (LO) phonon coupling strength, which showed reduction of coupling with decrease in well width. This reduction was explained in terms of confinement-induced enhancement of the excitonic binding energy by comparing the binding energy dependence of calculated coupling strength.
\end{abstract}
\enlargethispage*{2cm}

\maketitle
Blue or ultraviolet semiconducting light-emitting diodes have the possibility to revolutionize the future systems of display, illumination and information storage. While GaN (band-gap 3.5~eV at 2~K) and the related alloys have been emerging as the winner of this field, another wide-gap semiconductor, ZnO (3.4~eV) has also been attracting increasing interest because its excitons have a 60~meV binding energy as compared with 26~meV for GaN~\cite{LBZincoxide}. ZnO has additional advantageous properties: availability of large area substrates, higher energy radiation stability, environmentally friendly ingredients, and amenability to wet chemical etching. The main obstacle for the development of ZnO has been, however, the lack of reproducible p-type ZnO. Recently, the use of ``repeated temperature modulation epitaxy'' has given solution for this bottleneck~\cite{nature_mat_tsukazaki}. The electro-luminescence has been observed from homoepitaxial ZnO p-n junction structures. As stated in Ref.~\onlinecite{nature_mat_tsukazaki}, clear understanding of the physical processes of MgZnO alloys having a wider band-gap or ZnO/MgZnO heterostuctures is next challenge in order to control their material properties~\cite{makino29}. 

The excitonic features dominate the emission and absorption of ZnO-related heterostructures near the fundamental band gap. These features are always influenced by phonon-assisted processes, giving rise to the fact that an exciton line has finite width even at very low temperature. The optical absorption by interacting exciton-phonon system is thus a basic problem. Pelekanos \textit{et al.} reported significant change in the exciton-longitudinal-optical (LO) phonon interaction of ZnSe quantum wells (QW), sensitively dependent on QW width~\cite{pelekanos1}. This change has been attributed to the quantum confinement-related enhancement of excitonic binding energies with respect to the LO phonon energy~\cite{jpdoran1}, which is in contrast to III-V semiconductors such as GaAs~\cite{gammon1}. However, they have not experimentally determined their binding energies as a function of QW width quantitatively in that study. The stimulated emission characteristics of zinc oxide are dominated by exciton-exciton interaction process, which thus gives rise to an easy determination of the binding energies without the use of very high magnetic field.

In a previous short communication~\cite{sun2}, we presented experimental results on excitonic linewidths as well as their binding energies of ZnO QWs grown on ScAlMgO$_4$ substrates. A reduction of the excitonic-phonon coupling constant with a decrease in well layer thickness was observed.

In this work, linewidths decided by the interaction between excitons and LO phonons were calculated. It was found that this type of coupling is strongly dependent on QW width. A comparison of results of experiments and calculation was made. The experimental well-width dependence was reproduced well by calculation. The results of calculation, however, could not explain the difference in coupling strength between bulk ZnO and QWs. The calculated results also predicted resonant enhancement of the coupling around $\hbar \omega_{LO} = B_1$ ($B_1$ denoting the binding energy), which is in poor agreement with the experimental results.

Our samples were grown by laser molecular-beam
epitaxy~\cite{sun1,sun2,makino11}. We estimated the absorption linewidths of the lowest-lying excitons for
these structures as a function of measurement temperature. Their thermal broadening was thought to be due to exciton-phonon interaction. The full width at half maximum (FWHM) can be approximately described by the following equation:
\begin{equation}
    \Gamma(T) = \Gamma_{inh} + \gamma_{ph}T+ \Gamma_{LO}/[\exp(\hbar  \omega _{LO}/k _{B}T) - 1],
   \label{e2}
\end{equation}
where $\Gamma_{inh}$ is a temperature-independent term that denotes the inhomogeneous linewidth due to the exciton-exciton, exciton-carrier interactions~\cite{rudinandsegall1}, and the scattering by defects, impurities. The second term $\gamma_{ph} T$ is due to acoustic phonon scattering. This term represents the acoustic phonon coupling strength, mainly caused by a deformation potential mechanism. At low temperatures, the linewidth increases from the zero-temperature value
first because of the scattering by acoustic phonons. Note that the population of LO phonon is vanishingly small over this temperature range. The third term is the linewidth due to LO phonon scattering. $\Gamma_{LO}$ is the exciton-LO phonon coupling strength, and other symbols have the usual meanings. At high temperatures, the LO phonon Fr\"ohlich scattering dominates to the linewidth broadening. An LO-phonon energy of 72~meV was used.

\begin{figure}
\includegraphics[width=0.6\linewidth]{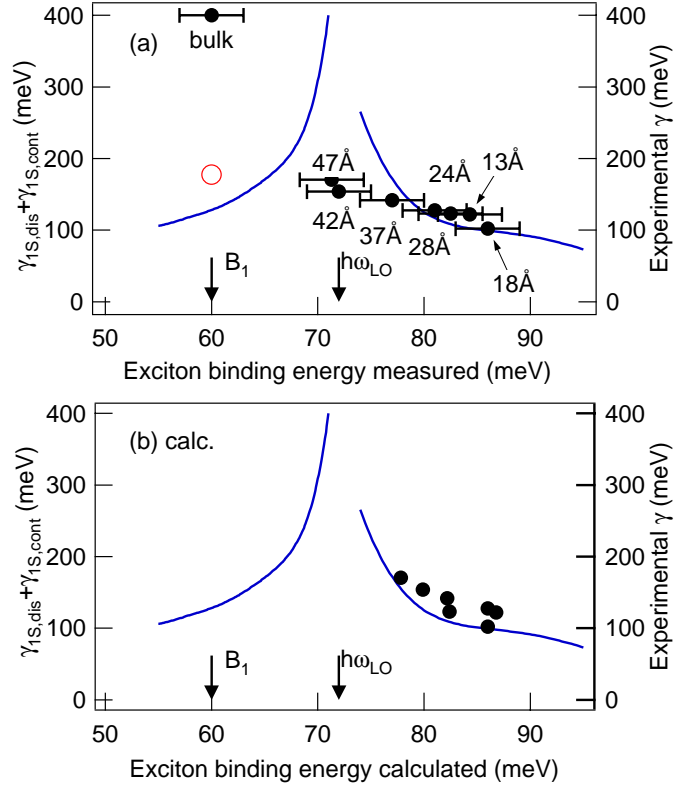}
\caption{(Color online): Experimental coupling strengths between excitons and LO phonons (full circles) plotted against the experimental (a) and calculated (b) exciton binding energies in ZnO bulk and quantum wells. Also shown by a solid line is the calculated curve for $\Gamma_{LO}$ based on Eqs.~(2) and (3). The well widths are marked next to the symbols in the upper frame. An open circle (datum for GaN) and results on calculated binding energies are cited from Refs.~\onlinecite{xbzhangGaNrev} and \onlinecite{coli1}, respectively.}
\label{fwhm}
\end{figure}
\begin{figure}
\includegraphics[width=0.35\linewidth]{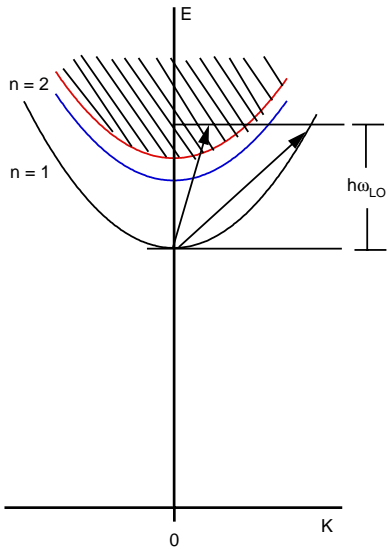}
\caption{(Color online): Schematic illustration of the scattering process contributing the phonon-induced broadening of the exciton linewidth. The solid lines indicate Rydberg series of exciton states, while the dashed line area depicts continuum states. \label{schematic}}
\end{figure}

From a fit to Eq.~(\ref{e2}), we determined the values for $\Gamma_{LO}$ which
were plotted for different QW widths in Fig.~\ref{fwhm} as a function of the exciton binding energy (closed circles). The experimental value of $\Gamma_{LO}$ of a ZnO epitaxial layer was also shown~\cite{makino8}. The experimental exciton binding energies ($B_1$) for QWs have been given in Ref.~\onlinecite{sunfull1}.  The plotted $\Gamma_{LO}$s are half of those given in our previous studies~\cite{makino8,sun2} because the values of $\Gamma_{LO}$ were previously determined by full width at half maximum (FWHM) values of experimental spectra. This has been done to accomodate with the theory which calculated its half width (HWHM). It should be noted that when $B_1$ is plotted against $L_w$ it has a relative maximum around
$L_w = 10 \textrm{\AA}$ due to the penetration of wavefunction ($L_w$ being QW width).
The exciton-phonon coupling in all of the QWs ($L_w \le 47$~\textrm{\AA}) assessed in this study was found to be smaller than that in bulk ZnO. In addition, the values of $\Gamma_{LO}$ monotonically decrease with a decrease in well width. We speculate that this variation can be explained in terms of the confinement-induced enhancement of excitonic binding energy. The processes involving LO-phonon interaction are shown in Fig.~\ref{schematic}, where the phonon can scatter the excitons to the same bound state, to higher-lying bound states, and to continuum states. For bulk ZnO, the binding energy is about 60~meV, smaller than that of the LO phonon (72~meV). Therefore, there must be many dissociation channels for the three-dimensional
excitons, giving rise to a larger $\Gamma_{LO}$. This is no longer
the case for the QWs. All of the wells have binding energies comparable with or greater than the LO phonon energy, which is constant over the entire range of well width as determined by the low-temperature PL spectra. In such a case,
dissociation efficiency of the discrete excitons into continuum states is suppressed.

To explain this tendency, we calculated contributions of both the bound and scatttered electron-hole states to the linewidth of the lowest-lying exciton using a model of Rudin \textit{et~al}. Since this model was established for bulk materials, it neglects effects of the phonon confinement as well as the formation of quantized states. We believe that such a simple model can reproduce
the main features of our experimental results. The details of the derivation can be found in Ref.~\onlinecite{rudinandsegall1}. Here we just give the final result, which for the discrete part of the spectrum is
\begin{equation}
\gamma_{1S, dis} = \hbar \omega_{LO} (R_0 \slash |E_0|) (\epsilon_\infty^{-1} - \epsilon_0^{-1}) \sum_{n =1}^{N} \Delta_n (q_1) |{\rm sgn} E_0 + n^{-2} B_1 \slash |E_0||^{-1/2},
\end{equation}
where $E_0 = \hbar \omega_{LO}$, $B_1$ is the binding energy of the ground-state exciton state, and $R_0$ is the Rydberg energy. $\delta_n$ is composed of a number of lengthy expressions that have been
given in the Appendix of Ref.~\onlinecite{rudinandsegall1}. The upper limit of this summation depends on whether the $\hbar \omega_{LO}$ is larger than $B_1$ or not. In the case of $\hbar \omega_{LO} > B_1$, the calculation should be summed up to $\infty$, while $N = (1- \hbar \omega_{LO} \slash B_1)^{-1/2}$ for the contrary case.

Next, the continuum part of the spectrum is discussed. As seen in Fig.~\ref{schematic}, transition to the electron-hole continuum or scattering states contribute to the linewidth if $\hbar \omega_{LO} > B_1$. This condition is satisfied in our case only for bulk. The finally obtained
expression for the linewidth parameter $\gamma_{1S, cont}$ is cited from Ref.~\onlinecite{rudinandsegall1} as follows:
\begin{equation}
\gamma_{1S, cont} = 128 \hbar \omega_{LO} (\epsilon_0 \slash \epsilon_{\infty} - 1) (M \slash \mu) \int_{0}^{x_0} dx \quad x (1+x) \left[ 1 - \exp (- 2 \pi \slash x) \right]^{-1} \left( F_{ee} + F_{hh} -2 F_{ee} \right),
\end{equation}

\begin{table}[htbp]
\begin{center}
\begin{ruledtabular}
\begin{tabular*}{\hsize}{l@{\extracolsep{0ptplus1fil}}c@{\extracolsep{0ptplus1fil}}r}
Name&Symbol&Value\\
\hline 
effective electron mass&$m_e^{*}$&0.24\\
effective hole mass&$m_h^{*}$&0.98\\
low-frequency dielectic contant&$\epsilon_{0}$&8.1\\
high-frequency dielectric contant&$\epsilon_{\infty}$&4.0\\
LO-phonon energy&$\hbar \omega_{LO}$&72~meV\\
excitonic binding energy&$B_1$&59~meV\\
\end{tabular*}
\end{ruledtabular}
\end{center}
	\caption{The material parameters of ZnO used in calculating the LO-phonon interaction to the linewidth, taken from the Landolt-B\"{o}rnstein table.}
\end{table}
We did not explicitely present $F_{ee}$, $F_{hh}$, and $F_{eh}$, which have somewhat lengthy expressions. These can also be found in the Appendix of Ref.~\onlinecite{rudinandsegall1}. $x_0 = (\mu \slash m_0)^{-1/2} \frac{1}{2} \epsilon_0 (E_0 \slash R_0)^{1/2}$, and the integration over $x$ was numerically performed. Here, $m_0$ is the vacuum electron mass and $R_0$ is 13.6~eV. The material parameters of ZnO are given in Table~I. It should be noted that the theory calculates the HWHM for $\Gamma_{LO}$. It is evidenced by the fact that, according to the calculation by Rudin \textit{et~al.}, the value for $\Gamma_{LO}$ of bulk ZnSe is 30~meV, while Pelekanos \textit{et~al.} experimentally estimated it to be at least 60~meV.

The calculated results $\gamma_{1S, dis} + \gamma_{1S, cont} = \Gamma_{LO}$ are shown in Fig.~\ref{fwhm}(a). On the QW side, the calculated $\Gamma_{LO}$ is a monotonically decreasing function of $B_1$, which is in qualitatively good agreement with the experimental results. On the other hand, calculated $\Gamma_{LO} $ exhibits a resonant enhancement behavior at $\hbar \omega_{LO} = B_1$ and appears to be greater in wide quantum wells than in a three-dimensional system, which does not agree with our experimental results. One of the reasons for this is probably experimental uncertainty of $B_1$ (typically $\pm 3$~meV), the error bars of which are shown in Fig.~\ref{fwhm}. Since $\hbar \omega_{LO}$ of wider QWs is very close to $B_1$, even a small change in $B_1$ leads to a huge variation in calculated $\Gamma_{LO}$. Coli and Bajaj~\cite{coli1} reported a calculated value of $B_1$ of 75~meV at a well width of 47~${\rm \AA }$. For comparison, a similar plot is shown in Fig.~\ref{fwhm}(b), where the abscissa now corresponds to calculated binding energies. The use of this value for the calculation of $\Gamma_{LO} $ yielded it of about 215~meV, which became closer to the experimental value. The calculation curve shows a sharp resonant enhancement structure at $B_1 = \hbar \omega_{LO}$. We speculate as a second reason that the theoretical calculation overestimated this resonance effect (too sharp) as can be understood by the fact that the final state of scattering events is not discrete but continuously spreaded in the spectrum. In addition, if the low dimensionality is taken into accout for QWs, the $\Gamma_{LO}$'s are expected to become smaller. The theory of Coli and Bajaj~\cite{coli1} has successfully described the experimental results on exciton binding energies and band gap energies of ZnO semiconductor quantum wells, but they have not yet given expression for exciton-phonon coupling strength. Since they have taken into account the interaction with phonon in their formulation so it may be possible to extend their theory so as to obtain the expressions like Eqs.~(2) and (3) of this paper, which is, however, beyond the scope of our work.

Finally, we comment on the very large value of experimental $\Gamma_{LO}$ in bulk ($\simeq 400$~meV in HWHM). There is large difference
between $\Gamma_{LO} $ in bulk and that in QWs. In the thin-film absorption spectra
obtained at $T \le 200$~K, peaks associated with the heavy-hole and light-hole excitons were spectrally resolved well, while this was
not the case for QWs, which probably leads to the difference in $\Gamma_{LO} $ values. Indeed, bulk value for $\Gamma_{LO}$ of ZnO
is exceptionally large compared with the values of 375~meV for GaN~\cite{xbzhangGaNrev} and 60~meV
for ZnSe (in FWHM). There is a large scattering in the experimental $\Gamma_{LO} $ values for
other semiconductors because of the variation in methods of determination. We used only data obtained by temperature-dependent
absorption spectroscopy. Since ZnO posesses an ionicity similar to that of
GaN, we can get better agreement of $\Gamma_{LO}$ with theory by assuming the value of 375~meV (187.5~meV in HWHM) to be a lower limit of $\Gamma_{LO}$ for ZnO bulk as shown by an open circle in Fig.~\ref{fwhm}. It might be more appropriate to deduce
$\Gamma_{LO} $ using a sufficiently wide QW ($L_w >> 100$~$\textrm{\AA}$)
in which the quantum size effect is not yet apparent so as to make $\Gamma_{inh}$ comparable each other.

In summary, we have studied the thermal broadening of exciton absorption linewidths in ZnO quantum wells. The exciton-LO-phonon coupling, the main cause of that broadening, is reduced in quantum well samples compared to that in bulk. This is explained in terms of a model in which low-dimensional confinement effects increases the exciton binding energy to a value greater than the LO-phonon energy and hence reduces the available wavenumber space for the exciton-LO-phonon scattering events.
A comparison with calculated results was made, and it was shown that the monotonical decrease in $\Gamma_{LO} $ with a decrease in well width is reproduced only qualitatively by the calculation. If the calculated results~\cite{coli1} on $B_1$ were used for $\Gamma_{LO}$ vs $B_1$ curve, a set of the experimental
$\Gamma_{LO}$ is in better agreement with the theoretical coupling parameter, $\gamma_{1S, \textrm{dis}}+\gamma_{1S, \textrm{cont}}$.

\textit{Acknowledgements} --- Authors thank A. Ohtomo and K. Tamura for sample preparations and N. T. Tuan, H. D. Sun and C. H. Chia for measurements. Thanks are also
due to H. Koinuma for continuous encouragement.


\end{document}